\documentstyle[12pt]{article}
\begin{document}
\thispagestyle{empty}
\begin{center}
\LARGE \tt \bf{Global monopoles and massless dilatons in Einstein-Cartan gravity}
\end{center}
\vspace{2.5cm}
\begin{center} {\large L.C. Garcia de Andrade\footnote{Departamento de
F\'{\i}sica Te\'{o}rica - Instituto de F\'{\i}sica - UERJ
Rua S\~{a}o Fco. Xavier 524, Rio de Janeiro, RJ
Maracan\~{a}, CEP:20550-003 , Brasil.
E-mail : GARCIA@SYMBCOMP.UERJ.BR}}
\end{center}
\vspace{2.0cm}
\begin{abstract}
A global monopole in dilatonic Einstein-Cartan gravity is presented.
A linearized solution representing a global monopole interacting 
with a massless dilaton is found where Cartan torsion does not 
interact with the monopole Higgs field.Computation of the geodesic 
equation shows that the monopole-dilaton system generates a repulsive gravitational field.The solution is shown to break the linear approximation for certain values of torsion.
\end{abstract}
\vspace{1.0cm}
\begin{center}
\large{PACS number(s) : 0420,0450}
\end{center}
\newpage
\pagestyle{myheadings}
\markright{\underline{Global monopoles and massless dilatons in Einstein-Cartan gravity.}}
Recently Garcia de Andrade and Barros \cite{1} have found that the 
presence of global monopoles in linearized Einstein-Cartan gravity 
\cite{2} lead to modified torsionic effects on physical quantities 
related to the global monopole such as the deficit angle and the 
density distribution which are very much in agreement with the 
predictions of COBE \cite{3}.Earlier Dando and Gregory \cite{4} 
have found the gravitational field of  a global monopole within the 
context of low energy string gravity ,where the global monopole 
couples with massive and massless dilatons.They found that for 
massless dilatons the spacetime is generically singular,whereas 
when the dilaton is massive the monopole induces a long range 
dilaton cloud.Since dilatons have been recently investigated in the 
context of Einstein-Cartan gravity \cite{5} it seems interesting to 
extend their investigated to spacetimes with torsion.Let us consider
the simplest model of Barriola-Vilenkin \cite{6} to build global 
monopoles which is given by the Lagrangean
\begin{equation}
L({\psi}^{i})=\frac{1}{2}{\nabla}_{a}{\psi}^{i}{\nabla}^{a}{\psi}^{i}-\frac{\lambda}{4}({\psi}^{i}{\psi}^{i}-{\eta}^{2})^{2}
\label{1}
\end{equation}
where ${\psi}^{i}$ is a triplet of real scalar fields,$i=1,2,3$ and
$a,b=0,1,2,3$ are spacetime indices.The model has a global $O(3)$ symmetry spontaneously broken to a global $U(1)$ symmetry by the vacuum choice $|{\psi}^{i}|={\eta}$.Static spherically symmetric solutiona are search describing a global monopole at rest.The field configuration for this monopole is written according to Dando and Gregory as ${\psi}^{i}={\eta}f(r)e^{i}$ where $e^{i}$ is the unit radial vector in the internal space of the monopole.We consider the metric of the monopole as written by
\begin{equation}
ds^{2}=B(r)dt^{2}-A(r)dr^{2}-r^{2}d{\Omega}^{2}
\label{2}
\end{equation}
where $d{\Omega}^{2}=d{\theta}^{2}+sin^{2}{\theta}d{\phi}^{2}$.
Since the linearized global monopole solution of Einstein-Cartan 
gravity in linear approximation has already been found we will not 
repeat it here, and go directly to the Global monopole in dilatonic 
Einstein-Cartan gravity.Its action is given by 
\begin{equation}
S={\int}d^{4}x(-g)^{\frac{1}{2}}[e^{-2{\phi}}(-R'-4({\nabla}{\phi})^{2}-V'({\phi}))+e^{2a{\phi}}L]
\label{3}
\end{equation}
where L is the monopole Lagrangean for the monopole as given in 
(\ref{1}),$V({\phi})$ is the potential for the dilaton.The action 
(\ref{3}) is written in terms of the string metric of the string 
sigma model.We write the action in terms of the conformal metric  
\begin{equation}
g_{ab}=e^{-2{\phi}}g'_{ab}
\label{4}
\end{equation}
which turns the action into
\begin{equation}
S={\int}d^{4}x(-g)^{\frac{1}{2}}[e^{-2{\phi}}(-R+2({\nabla}{\phi})^{2}-V({\phi}))+e^{2(a+2){\phi}}
L({\phi},e^{2{\phi}})]
\label{5}
\end{equation}
The energy-momentum tensor is decomposed in three parts.The first 
is the monopole one
\begin{equation}
T^{m}_{ab}=e^{-2{\phi}}{\nabla}_{a}{\phi}^{i}{\nabla}_{b}{\phi}^{i}-g_{ab}L
\label{6}
\end{equation}
the dilaton tensor 
\begin{equation}
S_{ab}=2{\nabla}_{a}{\phi}{\nabla}_{b}{\phi}+\frac{1}{2}V({\phi})-g_{ab}({\nabla}{\phi})^{2}
\label{7}
\end{equation}
and finally the torsion EMT 
\begin{equation}
T^{torsion}_{ab}=3S_{acd}S_{b}^{cd}-\frac{1}{2}g_{ab}S^{2}
\label{8}
\end{equation}
where $S^{2}=S_{abc}S^{abc}$ is the square of the torsion tensor 
$S_{abc}$ which $S_{023}$ is the only non-vanishing torsion 
component chosen along the radial direction.The Einstein-Cartan 
equation can then be written in the quasi-Einsteinian form
\begin{equation}
G_{ab}=\frac{1}{2}e^{2(a+2){\phi}}T^{m}_{ab}+S_{ab}+T^{torsion}_{ab}
\label {9}
\end{equation}
Although the contribution of torsion in the problem considered here 
seems minor we shall prove that the perturbation method will produce 
a new torsion dependent global monopole in the Einstein-Cartan 
dilatonic gravity.Since there is no direct interaction between 
torsion and the monopole or the dilatonic field the dilaton 
equation will be the same as the one considered by Dando and Gregory.Once more we take ${\eta}{\lambda}^{\frac{1}{2}}=1$ and the modified energy-momentum tensor is
\begin{equation}
T'^{t}_{t}={\alpha}({\gamma}(\frac{f'^{2}}{2A}+\frac{f^{2}}{r^{2}})+\frac{1}{4}{\delta})
\label{10}
\end{equation}
and
\begin{equation}
T'^{r}_{r}={\alpha}({\gamma}(-\frac{f'^{2}}{2A}+\frac{f^{2}}{r^{2}})+\frac{1}{4}{\delta})
\label{11}
\end{equation}
and
\begin{equation}
T'^{\theta}_{\theta}={\alpha}({\gamma}\frac{f'^{2}}{2A}+\frac{1}{4}{\delta})
\label{12}
\end{equation}
where ${\delta}=(f^{2}-1)$,${\alpha}=e^{2(a+2){\phi}}$ and ${\gamma}=e^{-2{\phi}}$.The $tt$ and $rr$ components of Einstein-Cartan-dilatonic-global monopole system now becomes 
\begin{equation}
\frac{A'}{rA^{2}}+\frac{1}{r^{2}}(1-\frac{1}{A})={\epsilon}T'^{t}_{t}+\frac{1}{2}V({\phi})+\frac{{{\phi}'}^{2}}{A}+\frac{b}{r^{2}}
\label{13}
\end{equation}
and
\begin{equation}
\frac{-B'}{rB^{2}A^{2}}+\frac{1}{r^{2}}(1-\frac{1}{A})={\epsilon}T'^{r}_{r}+\frac{1}{2}V({\phi})-\frac{{{\phi}'}^{2}}{A}+\frac{b}{r^{2}}
\label{14}
\end{equation}
where we have chosen the Barros ansatz for torsion of the global 
monopole as $S^{2}=\frac{(1-b){\eta}^{2}}{r^{2}}$ where $b$ is a 
constant which when zero reduces the solution to the Barriola-
Vilenkin solution in the absence of dilatons.To simplify matters we 
consider in this Letter just massless dilatons leaving the massive 
dilatons case for a future paper.For massless dilatonic Einstein-
Cartan gravity $V({\phi})=0$,where ${\epsilon}=\frac{{\eta}^{2}}{2}$
and making the expansion
\begin{equation}
A=1+{\epsilon}A_{1}+...
\label{15}
\end{equation}
and
\begin{equation}
B=1+{\epsilon}B_{1}+...
\label{16}
\end{equation}
and
\begin{equation}
{\phi}={\phi}_{0}+{\epsilon}{\phi}_{1}+...
\label{17}
\end{equation}
To the first order we have,$A'=B'=0$ and the equations reduce to
\begin{equation}
{{\phi}'_{0}}^{2}-\frac{b}{r^{2}}=0
\label{18}
\end{equation}
which yields the simple solution ${{\phi}_{0}}^{2}={b}^{\frac{1}{2}}lnr$.Substitution of these values 
into fields equations we obtain
\begin{equation}
\frac{A'_{1}}{r}+\frac{A_{1}}{r^{2}}={\beta}r(\frac{{f'}^{2}}{2}+\frac{f^{2}}{r^{2}})+\frac{1}{4}{\beta}r(f^{2}-1)^{2}+\frac{b}{r^{2}}(A_{1}-1)
\label{19}
\end{equation}
and
\begin{equation}
\frac{-B'_{1}}{r}+\frac{A_{1}}{r^{2}}={\beta}r(-\frac{{f'}^{2}}{2}+\frac{f^{2}}{r^{2}})+\frac{1}{4}{\beta}r(f^{2}-1)^{2}+\frac{b}{r^{2}}(A_{1}+1)
\label{20}
\end{equation}
\begin{equation}
{{\phi}_{1}}"+\frac{2{{\phi}_{1}}'}{r}=(a+1)\frac{\beta}{r}(-\frac{{f'}^{2}}{2})+\frac{a+2}{4}(f^{2}-1)^{2}
\label{21}
\end{equation}
For large r,$f$ is approximatly unity which greatly simplifies the 
system to
\begin{equation}
{{\phi}_{1}}"+\frac{2{{\phi}_{1}}'}{r}=0
\label{22}
\end{equation}
which can be solved to yield
\begin{equation}
{{\phi}_{1}}'=er-\frac{2c}{{r}^{3}}+d
\label{23}
\end{equation}
where here we are using ${\beta}=e^{{2(a+1)}{\phi}_{0}}$.The field 
equations for the metric are
\begin{equation}
\frac{A'_{1}}{r}+\frac{A_{1}}{r^{2}}=\frac{\beta}{r}+\frac{b}{r^{2}}A_{1}
\label{24}
\end{equation}
and
\begin{equation}
\frac{-B'_{1}}{r}+\frac{A_{1}}{r^{2}}=\frac{\beta}{r}+\frac{b}{r^{2}}A_{1}
\label{25}
\end{equation}
Their solution is
\begin{equation}
A_{1}+B_{1}=0
\label{26}
\end{equation}
Substitution of the solutions of these equations on the expansion 
we obtain the following final metric
\begin{equation}
ds^{2}=(1-\frac{{\epsilon}{\beta}r}{2(b-1)}-{\epsilon}\frac{m}{r})dt^{2}-(1+\frac{{\epsilon}{\beta}r}{2(b-1)}+{\epsilon}\frac{m}{r})dr^{2}-r^{2}d{\Omega}^{2}
\label{27}
\end{equation}
where $m$ is an integration constant.This implies the following geodesics
\begin{equation}
\dot{v_{r}}=\frac{{\beta}}{2(b-1)}+\frac{m}{r^{2}}
\label{28}
\end{equation}
which very far from the sources shows that the dilaton-global 
monopole-torsion system is gravitationally repulsive.Note that the torsion factor b is determinant for the repulsive carachter of the global monopole-dilaton system.Notice that due to the presence of ${\beta}r$.Therefore the spacetinme should be bound to values of $r<(b-1)$ and the need for an exact solution is of prime importance for the better understanding of the problem.A more detailed account of the ideas discussed here may appear elsewhere.
\section*{Acknowledgements}
I am very much indebt to Prof.Shapiro and Prof.R.Ramos for helpful discussions on the subject of this paper. Financial support from CNPq. and UERJ (FAPERJ) is gratefully acknowledged.

\end{document}